\documentclass[12pt,dvips]{article}

\usepackage{rotating}
\usepackage{axodraw} 
\usepackage{epsfig}  
\usepackage{epsf}    
\usepackage{psfig}   
\usepackage{cite}    

\newcommand{\imag}{\Im {\rm m}}
\newcommand{\real}{\Re {\rm e}}

\def\lsim{\stackrel{<}{{}_\sim}}
\def\gsim{\stackrel{>}{{}_\sim}}

\begin{document}

\begin{flushright}
CERN-PH-TH/2005-031 \\
hep-ph/0502251 \\
February 2005
\end{flushright}

\begin{center}
{\bf {\LARGE Diffraction as a CP and Lineshape Analyzer}}\\[3mm]
{\bf {\LARGE for MSSM Higgs Bosons at the LHC }}
\end{center}

\bigskip

\begin{center}
{\large John Ellis$^{\,a}$, Jae Sik Lee$^{\,b}$
                                       and Apostolos Pilaftsis$^{\,b}$}
\end{center}

\begin{center}
{\em $^a$Theory Division, Physics Department, CERN, CH-1211 Geneva 23,
Switzerland}\\[2mm]
{\em $^b$School of Physics and Astronomy, University of Manchester}\\
{\em Manchester M13 9PL, United Kingdom}
\end{center}

\bigskip\bigskip\bigskip

\centerline{\bf ABSTRACT}
\noindent
We study the production and decay of a coupled system of mixed neutral
MSSM  Higgs bosons  in exclusive  double-diffractive processes  at the
LHC,    including    non-vanishing    CP    phases   in    the    soft
supersymmetry-breaking  gaugino masses and  third-generation trilinear
squark couplings. The three  neutral Higgs bosons are naturally nearly
degenerate, for  large values of  $\tan\beta$, when the  charged Higgs
boson weighs around  150 GeV.  Large mixing between  all three neutral
Higgs  bosons is  possible when  CP  is violated,  a three-way  mixing
scenario which  we also  term tri-mixing.  A  resolution in  the Higgs
mass of $\sim  1$~GeV, which may be achievable  using the missing-mass
method, would allow one  to distinguish nearly degenerate Higgs bosons
by   studying   the  production   lineshape.    Measurements  of   the
polarizations of  the tau leptons  coming from the  Higgs-boson decays
could  offer a direct  and observable  signal of  CP violation  in the
Higgs sector.

\newpage

\section{Introduction}
\label{sec:introduction}

Direct search experiments at LEP established that, within the Standard
Model,  the single  physical Higgs  boson must  weigh more  than about
$114$~GeV~\cite{LEPHWG},  and  the  mass  range favoured  by  indirect
measurements is  close to  this lower limit~\cite{LEPEWWG}.   Both the
direct and  indirect limits must  be re-examined in  non-minimal Higgs
scenarios.   One   popular  example  is   the  minimal  supersymmetric
extension of the Standard  Model (MSSM)~\cite{HPN}, which predicts the
existence of at least one light neutral Higgs boson weighing less than
about $135$~GeV.  This may be  joined by two other light neutral Higgs
bosons if the  charged Higgs bosons $H^\pm$ are also  light.  If CP is
conserved, only  the two  CP-even neutral Higgs  bosons $h,H$  may mix
with   each  other   dynamically,   through  off-diagonal   absorptive
self-energy effects. Instead, if CP is violated, all the three neutral
Higgs  bosons,  including the  CP-odd  neutral  Higgs  boson $A$,  mix
through CP-violating  quantum effects~\cite{APLB,PW,Demir,CDL,CEPW} to
mass eigenstates, $H_{1,2,3}$, of  indefinite CP. These CP-mixed Higgs
states  give rise  to  a  coupled system  whose  resonant dynamics  is
properly  described by  a 3-by-3  propagator matrix~\cite{APNPB,ELP1}.
One very interesting phenomenological feature of the CP-violating MSSM
is that the lightest neutral Higgs boson could be considerably lighter
than 114~GeV~\cite{CPX,CEMPW}.

Inclusive experiments at the LHC will be able to discover the Standard
Model Higgs  boson, whatever  its mass.  They  should also be  able to
discover the lightest  neutral MSSM Higgs boson --- at  least if CP is
conserved --- and will be  able to explore significant mass ranges for
the  heavier  MSSM Higgs  bosons.   Supplementing inclusive  searches,
interest has recently  been growing in the search  for Higgs bosons in
diffractive events at the LHC~\cite{DeRoeck,diffLHC}.  These may offer
novel prospects  for measuring the  properties of light  neutral Higgs
bosons~\cite{CFLMP}, and disentangling their CP properties~\cite{KMR}.

The MSSM offers  additional sources of CP  violation beyond the single
Kobayashi--Maskawa phase      in the Standard    Model.   If  the soft
supersymmetry-breaking  parameters   $m_0,   m_{1/2}$   and    $A$ are
universal, two new physical  CP-odd phases remain. These may,  without
loss   of generality, be  parametrized  as one  phase in the trilinear
couplings $A$ and one in the  gaugino masses $m_{1/2}$. In addition to
signatures   of  CP violation in   sparticle   production and decay at
high-energy   colliders~\cite{CPdirect,CPsoft1,CPsoft2},  these phases
may have observable radiative effects on the Higgs sector~\cite{APLB},
on   electric  dipole    moments~\cite{EDM1,EDM2,CKP},  and  in    $B$
decays~\cite{Bmeson1,DP}.  One   of  the  principal   motivations  for
studying such models is  the prospect  of electroweak baryogenesis  in
the MSSM~\cite{baryog}.

In this  paper, we consider  the prospects for studying  light neutral
MSSM Higgs  bosons in diffractive  events at the LHC,  particularly in
scenarios                  where                 CP                 is
violated~\cite{APLB,PW,Demir,CDL,CEPW,INhiggs,KW,HeinCP,CEPW2,Maria},
and the three neutral Higgs  bosons mix strongly.  This work continues
previous studies  of the masses,  couplings, production and  decays of
the  mixed-CP Higgs  bosons $H_{1,2,3}$,  with a  view to  searches at
LEP~\cite{CPX},  the  LHC~\cite{CHL,CPX,CEMPW,CPpp,CFLMP,Akeroyd,KMR},
the  ILC~\cite{CPee},  a  $\mu^+\mu^-$  collider~\cite{CPmumu}  and  a
$\gamma \gamma$ collider~\cite{CPphoton,PPTT,GKS,CKLZ,ELP2}. As in our
previous  works~\cite{ELP1,ELP2}, we include  a complete  treatment of
loop-induced CP violation and three-way mixing, including off-diagonal
absorptive   effects   in    the   resummed   Higgs-boson   propagator
matrix~\cite{APNPB}.

Higgs-boson   production in an   exclusive  diffractive collision $p+p
\rightarrow p+H_i+p$, where  the  outgoing protons remain intact   and
scatter   through  small  angles,  offers  a  unique  environment  for
investigating the MSSM Higgs sector, in particular when $\tan\beta$ is
large and $M_{H^{\pm}}^{\rm pole}\sim 150$ GeV. In such scenarios, all
three neutral  Higgs bosons have   similar masses and there  is strong
three-way mixing.   We call a scenario  with these properties the {\em
three-way mixing} or {\em tri-mixing} scenario of the CP-violating MSSM.

Moreover, the production cross section of  the Higgs boson can be much
enhanced in the MSSM for large values of $\tan\beta$, as compared with
the Standard Model.  Furthermore,  good Higgs-mass resolution  of  the
order of 1 GeV may  be achievable~\cite{DeRoeck} by precise measurements
of the momenta  of  the  outgoing  protons in   detectors a long   way
downstream of  the interaction point. This  enables one to disentangle
nearly degenerate Higgs bosons  by examining the  production lineshape
of the coupled system of neutral Higgs bosons.

The layout  of this  paper  is  as  follows. Section~2  provides basic
formulae   for the  luminosity  of   the exclusive  double-diffractive
process, based on~\cite{KMRLUM,JEFF}. Section~3 presents the formalism
for the  production,    mixing  and decay  of   a  coupled   system of
CP-violating     neutral MSSM Higgs  bosons     in diffraction,  based
on~\cite{ELP1}.  It  also presents  numerical example  in a  couple of
CP-violating scenarios, considering two final states: ${\bar b} b$ and
$\tau^+ \tau^-$. Our conclusions are given in Section~4.

\section{Luminosity for the Exclusive Double Diffractive\\ Process}
\label{sec:luminosity}

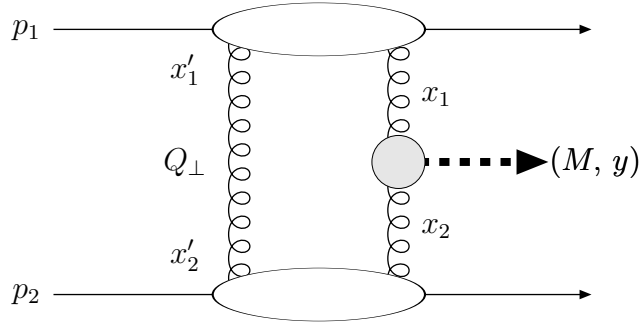
\begin{figure}[tbh]
\vspace*{1cm}
\begin{center}
\begin{picture}(300,100)(0,0)

\ArrowLine(50,0)(250,0)\ArrowLine(250,0)(251,0)
\ArrowLine(50,100)(250,100)\ArrowLine(250,100)(251,100)
\Gluon(120,0)(120,100){4}{10}
\Gluon(180,0)(180,100){4}{10}
\SetWidth{3}
\DashLine(180,50)(230,50){4}\ArrowLine(230,50)(231,50)
\SetWidth{0.2}
\GOval(150,0)(10,40)(0){1.0}
\GOval(150,100)(10,40)(0){1.0}
\GOval(180,50)(10,10)(0){0.9}
\SetWidth{0.3}

\Text(40,100)[c]{$p_1$}\Text(40,0)[c]{$p_2$}
\Text(100,85)[c]{$x^\prime_1$}\Text(100,15)[c]{$x^\prime_2$}
\Text(100,50)[c]{$Q_{\perp}$}
\Text(195,75)[c]{$x_1$}\Text(195,25)[c]{$x_2$}
\Text(255,50)[c]{$(M,\,y)$}
\Text(255,50)[c]{$(M,\,y)$}

\end{picture}\\
\end{center}
\smallskip
\noindent
\caption{\it The mechanism contributing to
the exclusive double-diffractive process.
}
\label{fig:diffraction}
\end{figure}

The effective luminosity for producing via double diffraction a system of 
invariant mass $M$ and rapidity $y$ can be written as \cite{KMRLUM}
\begin{equation}
  \label{effL}
M^2\frac{\partial^2 {\cal L}}{\partial y\partial M^2}\ =\ \hat{S}^2 L\;,
\end{equation}
where   the  `soft' survival factor  $\hat{S}^2$  is  quite model- and
process-dependent.  Denoting the longitudinal   momenta of  two gluons
that fuse into a system with invariant mass $M$ by  $x_1 p_1$ and $x_2
p_2$, we have
\begin{equation}
x_1\ =\ \frac{M}{\sqrt{s}}\; e^{y}\,,  \qquad\qquad
x_2\ =\ \frac{M}{\sqrt{s}}\; e^{-y}\,,
\end{equation}
where $s \equiv (p_1+p_2)^2$ is the collider centre-of-mass energy 
squared, see Fig.~\ref{fig:diffraction}.

Assuming  $J_z=0$, a parity-even  central  system and negligibly small
perpendicular momenta  of the outgoing  protons: $p_\perp \ll  1$ GeV,
the expression for $L$ becomes~\cite{KMRLUM,JEFF}
\begin{equation}
L\ =\ \left[\frac{\pi}{(N_C^2-1)b}
\int_{Q_{\rm min}^2}^{\mu^2}\frac{{\rm d}Q_T^2}{Q_T^4}\,
f_g(x_1,x_1^\prime,Q_T^2,\mu^2)\,
f_g(x_2,x_2^\prime,Q_T^2,\mu^2)\right]^2\,,
\end{equation}
where $N_C=3$ and  $Q_T^2$ is the virtuality of  the soft gluon needed
for  colour  screening.   The  hard  scale  $\mu$  and  the  $t$-slope
parameters  are chosen  as  follows: $\mu=M/2$  and $b=4$  GeV$^{-2}$.
Formally, we also have  introduced a non-vanishing cutoff $Q_{\rm min}
\stackrel{<}{{}_\sim} 1$  GeV to  avoid encountering the  Landau pole,
but the sensitivity of $L$ to $Q_{\rm min}$ turns out to be not large.
Furthermore, if $x^\prime  \ll x$ (which is actually  the case for the
exclusive  diffractive   process~\cite{KMRLUM,JEFF}),  the  skewed  or
off-diagonal unintegrated  gluon density $f_g(x,x^\prime,Q_T^2,\mu^2)$
may take on the factorizable form:
\begin{equation}
f_g(x,x^\prime,Q_T^2,\mu^2)\ \simeq\ R_g\,
\tilde{f}_g(x,Q_T^2,\mu^2)\; ,
\label{eq:fgg}
\end{equation}
where  $R_g$ is  a  constant.  The  simplified form~(\ref{eq:fgg})  is
estimated  to have  an  accuracy of  10  to 20\%,  where the  function
$\tilde{f}_g(x,Q_T^2,\mu^2)$ is defined by
\begin{eqnarray}
\tilde{f}_g(x,Q_T^2,\mu^2) &\equiv & \frac{\partial}{\partial \ln Q_T^2}
\left[\sqrt{T(Q_T,\mu)}\,x\,g(x,Q_T^2)\right] \nonumber \\
&=& \frac{1}{2} \sqrt{T(Q_T,\mu)} x\,Q_T
\left[\frac{{\rm d}g(x,Q_T^2)}{{\rm d}Q_T}-
\frac{g(x,Q_T^2)}{2}\,\frac{{\rm d}S(Q_T,\mu)}{{\rm d}Q_T}\right]\,.
\label{eq:fg}
\end{eqnarray}
We note  that $\tilde{f}_g(x,Q_T^2,\mu^2)$ consists  of derivatives of
the gluon  distribution function  $g(x,Q_T^2)$ and the  Sudakov factor
$T(Q_T,\mu)  =  e^{-S(Q_T,\mu)}$.   Collecting  all the  factors,  the
effective luminosity can be rewritten as
\begin{equation}
M^2\frac{\partial^2{\cal L}}{\partial y\partial M^2}\ =\ \hat{S}^2
\left[\frac{\pi\,R_g^2}{8\,b}
\int_{\ln\!{Q_{\rm min}}}^{\ln\!{\mu}}
F_g(x_1,x_2,Q_T,\mu)\,{\rm d}\ln\!{Q_T} \right]^2\;,
\end{equation}
where the integrand functions $F_g(x_1,x_2,Q_T,\mu)$ is defined by
\begin{equation}
  \label{eq:lum}
F_g(x_1,x_2,Q_T,\mu)\ \equiv\  2\;\frac{\tilde{f}_g(x_1,Q_T^2,\mu^2)
\tilde{f}_g(x_2,Q_T^2,\mu^2)}{Q_T^2}\ ,
\end{equation}
for the double-diffractive process.

For    our  numerical analysis,    we also  need  to   know the parton
distribution       functions      (PDFs)     for   $g(x,Q_T^2)$     in
$\tilde{f}_g(x,Q_T^2,\mu^2)$.  Specifically, we take the PDFs given by
CTEQ6M~\cite{CTEQ6} and  MRST2004NNLO~\cite{MRST2004}.    The function
$S(Q_T,\mu)$ that determines the  Sudakov factor $T (Q_T,\mu)$ can  be
calculated as
\begin{equation}
S(Q_T,\mu)\ =\ \frac{1}{2\pi}\int_{Q_T^2}^{\mu^2}
\frac{\alpha_s(k_t^2)}{k_t^2}
\left[{\cal F}_g(k_t)\ +\!\!\sum_{q=u,d,s,c,\bar{u},\bar{d},\bar{s},\bar{c}}
{\cal F}_q(k_t)\right] {\rm d}k_t^2\,,
\end{equation}
where
\begin{eqnarray}
  \label{Fgq}
{\cal F}_g(k_t)&=& \int_0^{1-\Delta(k_t)} z P_{gg}(z) \,{\rm d}z\
=\ -11/2-6\ln\Delta+12\Delta-9\Delta^2+4\Delta^3-3\Delta^4/2\;,
\nonumber \\
{\cal F}_q(k_t)&=& \int_0^{1-\Delta(k_t)} P_{qg}(z) \,{\rm d}z\
=\ 1/3-\Delta/2+\Delta^2/2+\Delta^3/3\;,
\end{eqnarray}
with   $\Delta=\Delta(k_t)\equiv   k_t/(\mu+k_t)$.

\begin{figure}[htb]
\vspace{-1.5cm}
\centerline{\epsfig{figure=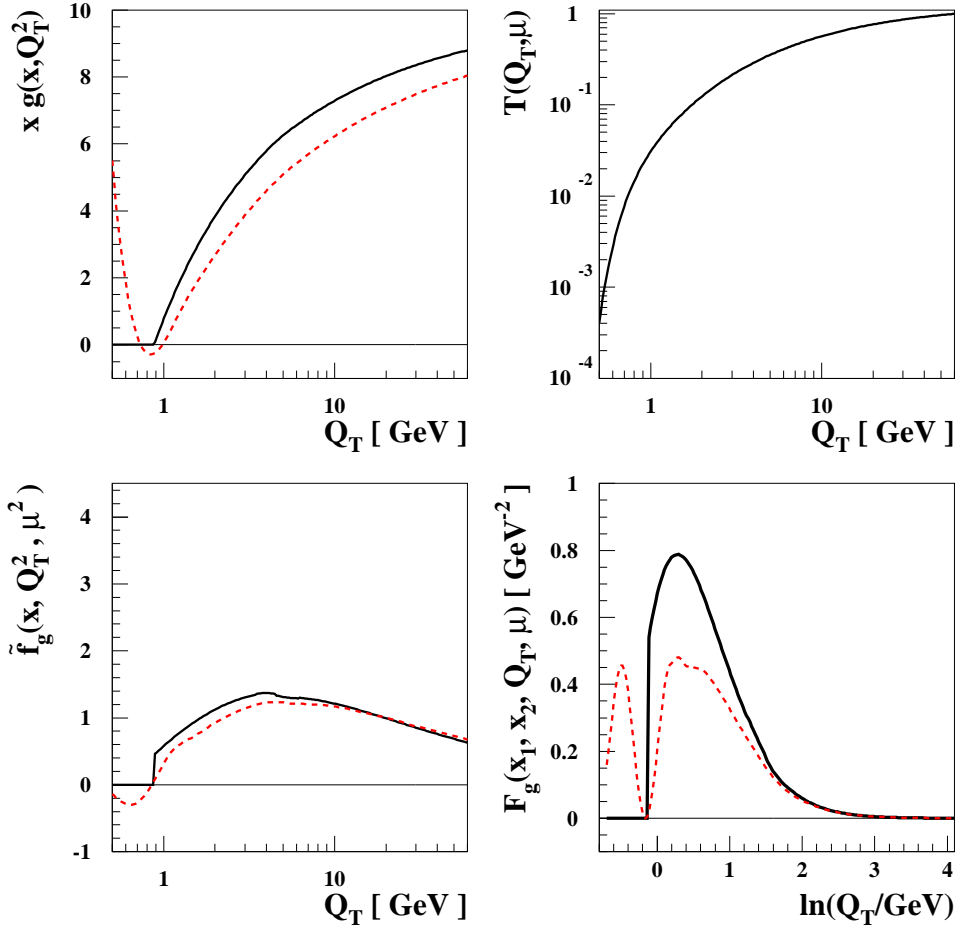,height=14cm,width=14cm}}
\vspace{-0.9 cm}
\caption{{\it 
The quantities $xg(x,Q_T^2)$ (upper left), the Sudakov factor
$T(Q_T,\mu)$ (upper right), $\tilde{f}_g(x,Q_T^2,\mu^2)$ (lower left), 
and $F_g(x_1,x_2,Q_T,\mu)$ (lower right),
as functions of $Q_T$ or $\ln{Q_T}$ when the
rapidity $y=0$, $M=120$ GeV, and $\sqrt{s}=14$ TeV. The solid (dashed)
lines are the outputs of CTEQ6M (MRST2004NNLO).
}}
\label{fig:fgm120y0}
\end{figure}

Equations (\ref{effL})--(\ref{Fgq}) provide  a complete basis  for our
numerical   evaluation  of   the  effective  luminosity  for exclusive
double-diffractive Higgs production.   In particular, we find that the
effective  luminosity may conveniently be computed as follows:
\begin{equation}
M^2\frac{\partial^2{\cal L}}{\partial y\partial M^2}\ =\ 4.0\times 10^{-4}
\left[ \frac{\int_{\ln\!{Q_{\rm min}}}^{\ln\!{\mu}}
F_g(x_1,x_2,Q_T,\mu)\,{\rm d}\!\ln\!{Q_T}}{\rm GeV^{-2}} \right]^2
\left( \frac{\hat{S}^2}{0.02}    \right)
\left( \frac{4}{b~{\rm GeV}^2}\right)^2
\left( \frac{R_g}{1.2}           \right)^4\; .
\label{eq:numlum}
\end{equation}
We note that the luminosity is very sensitive  to the choice of $R_g$, in
particular, but less to the hard scale $\mu$ (see also our
discussion below).

In Fig.~\ref{fig:fgm120y0}, we show $xg(x,Q_T^2)$ (upper left), 
the Sudakov factor
$T(Q_T,\mu)$ (upper right), $\tilde{f}_g(x,Q_T^2,\mu^2)$ (lower left), 
and $F_g(x_1,x_2,Q_T)$ (lower right), see (\ref{eq:lum}),
as functions of $Q_T$ or $\ln{Q_T}$ when the rapidity $y=0$, $M=120$ GeV, 
and $\sqrt{s}=14$ TeV. The solid (dashed) lines are outputs of CTEQ6M 
(MRST2004NNLO).
We observe significant differences between the functions $xg(x,Q_T^2)$,
which results in a strong dependence on the parton distribution function
used. Specifically, MRST2004 gives a non-vanishing and partly negative 
$xg(x,Q_T^2)$ when $Q_T\lsim 0.9$ GeV, where 
$\tilde{f}_g(x_{1,2},Q_T^2,\mu^2)$ becomes negative
due to the rapidly decreasing $xg(x,Q_T^2)$. On the other hand, CTEQ6M returns
$g(x,Q_T^2)=0$ in this region.
For $0.9~{\rm GeV} < Q_T < 10$ GeV, MRST2004 returns smaller values for
$F_g(x_1,x_2,Q_T,\mu)$ than CTEQ6M does.
When $\ln(Q_T/{\rm GeV}) > 3$  ($Q_T \gsim 20~{\rm GeV}$), 
the contribution from this large-$Q_T$ region to the luminosity is 
negligible,
making the prediction insensitive to the choice of $\mu \gsim 20$ GeV.
We note that the most significant contribution comes from the region 
around $Q_T=1.3$ GeV, as seen in the lower-right frame for 
$F_g(x_1,x_2,Q_T,\mu)$.

In fact, the calculations of $g(x,Q_T^2)$ in both the CTEQ6 and MRST2004
parameterizations are not reliable when $Q_T\lsim 1$ GeV. There are some
phenomenological prescriptions for making the low-$Q_T^2$ behaviour more
sensible~\cite{LOWQ2}.  For Higgs bosons with masses $M \sim 120$~GeV,
however, the luminosity does not depend strongly on the variation between
the prescriptions, and we take $Q_{\rm min}=1$ GeV to avoid unphysical
effects associated with the inapplicability of the $g(x,Q_T^2)$
calculation for $Q_T\lsim 1$ GeV.

\begin{figure}[htb]
\vspace{-1.5cm}
\centerline{\epsfig{figure=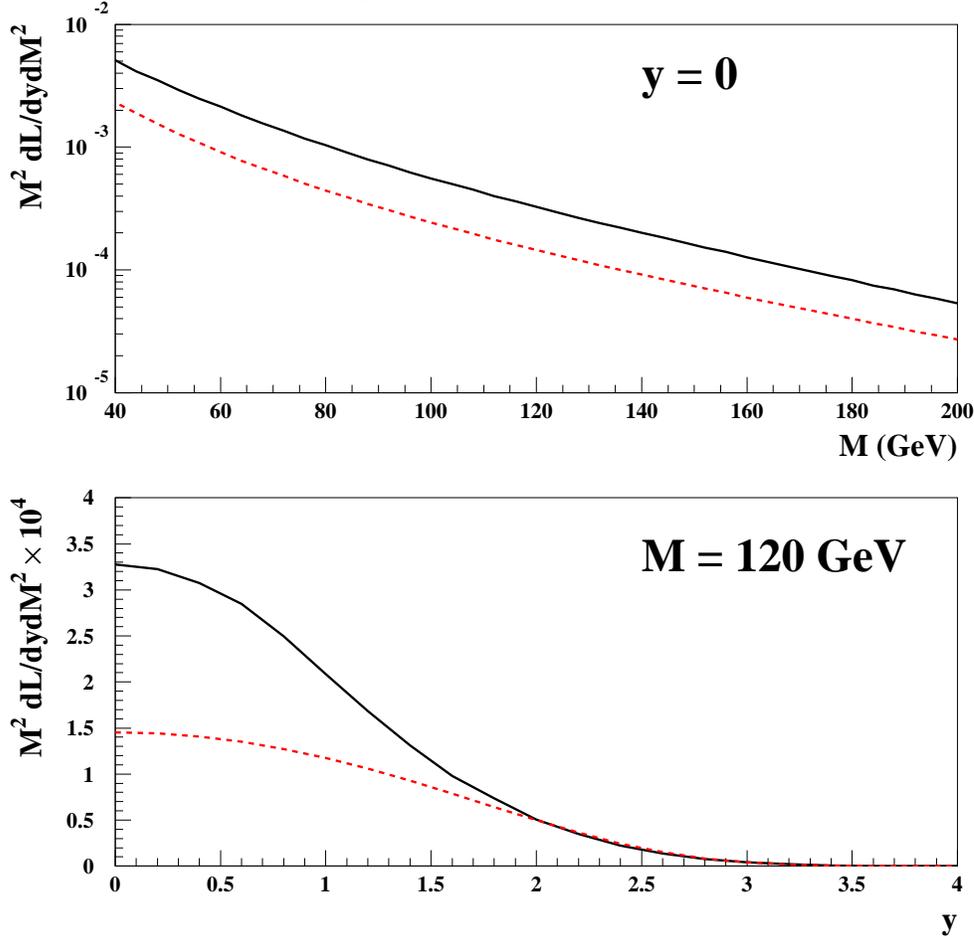,height=14cm,width=14cm}}
\vspace{-0.9 cm}
\caption{{\it The effective luminosity for the exclusive double-diffractive
process for $\sqrt{s}=14$ TeV as a function of $M$ ($y=0$) and $y$  
($M=120$~GeV) in the upper (lower) frame, respectively. Solid (dashed) 
lines are obtained using CTEQ6M (MRST2004NNLO), taking $Q_{\rm min}=1.0$ 
GeV in each case. Predictions using other PDFs lie between the above
two extreme cases, according to the analysis in~\cite{JEFF}.}}
\label{fig:lum}
\end{figure}

In  Fig.~\ref{fig:lum},  we  show  the effective  luminosity  for  the
exclusive double-diffractive process  at the LHC as a  function of $M$
for  $y=0$ (upper  frame) and  as a  function of  $y$ for  $M=120$ GeV
(lower).  The  solid (black) lines  are obtained using CTEQ6M  and the
dashed  (red) lines by  MRST2004NNLO. We  take $Q_{\rm  min}=1.0$ GeV,
$\hat{S}^2=0.02$, and $R_g=1.2$~\cite{KMRLUM}, see~(\ref{eq:numlum}).
We  note large  differences  between the  two  predictions over  large
regions of $M$ and $y$, which should be regarded as two extreme cases.
For example,  at $(y,M)=(0,120~{\rm  GeV})$, the CTEQ6M  prediction is
about  twice as  large  as that  of  MRST2004NNLO, and we find that the
prediction of CTEQ6M  has a stronger dependence on  $y$ than does that
of  MRST2004NNLO. We  remark  that  the
predictions using other PDFs lie  between the above two extreme cases,
as discussed in~\cite{JEFF}.
%

\section{The Process $pp\rightarrow p+H_i+p\rightarrow
p+[f(\sigma)\bar{f}(\bar{\sigma})]+p$}
\label{sec:formalism}

The         helicity     amplitude         for       the       process
$g^a_1(\lambda_1)g^b_2(\lambda_2)\rightarrow      H_i      \rightarrow
f(\sigma)\bar{f}(\bar\sigma)$  in the double-diffractive production of
Higgs bosons $H_i$ is given by
\begin{equation}
{\cal M}(\sigma\bar\sigma;\lambda_1\lambda_2)\ =\ 
\frac{g\alpha_sm_f\sqrt{\hat{s}}\delta^{ab}}{8\pi v M_W}\;
\langle\sigma\rangle_f\;
\delta_{\sigma\bar\sigma}\delta_{\lambda_1\lambda_2}\; ,
\end{equation}
where the amplitude $\langle\sigma\rangle_f$ is defined as
\begin{equation}
\langle\sigma\rangle_f\ \equiv\
\sum_{i,j=1,2,3} S_i^g(\sqrt{\hat{s}})\, 
D_{ij}(\hat{s})\,(\sigma\beta_f g^S_{H_j\bar{f}f} -ig^P_{H_j\bar{f}f})\;.
\end{equation}
For the $J_z^P=0^+$ process with $p_\perp \approx 0$ that we consider,
the  pseudoscalar form  factor of  the $g$-$g$-$H_i$  vertex  does not
contribute,   making  the  helicity   amplitude  independent   of  the
helicities  of  gluons. For  the  definitions  of  the couplings,  the
threshold  corrections   that  are   enhanced  for  large   values  of
$\tan\beta$  for $f=b\,,\tau$,  and  the full  $3\times 3$  propagator
matrix $D_{ij}(\hat{s})$, we refer to~\cite{ELP1,CPsuperH}.

Similarly as in~\cite{ELP1}, one can define the parton-level cross section 
as:
\begin{equation}
\hat\sigma_i^f\ \equiv\ 2 (N_C^2-1) \frac{N_f\beta_f}{512\pi\hat{s}}
\left(\frac{g\alpha_sm_f\sqrt{\hat{s}}}{8\pi v M_W}\right)^2 C^f_i
\end{equation}
where the enhancement factor $2(N_C^2-1)$ for the exclusive process
and color factors $N_f: N_l=1$ and $N_q=3$ have been included.  
The coefficients $C_i^f$ are given in terms of the amplitudes
$\langle\sigma\rangle_f$:
\begin{eqnarray}
C^f_1\!&=&\! \frac{1}{2} (|\langle + \rangle_f|^2 + |\langle -
\rangle_f|^2)\,, \qquad
C^f_2\ =\ \frac{1}{2} (|\langle + \rangle_f|^2 - |\langle - \rangle_f|^2)\,, 
\nonumber \\
C^f_3 \!&=&\! - \real(\langle + \rangle_f \langle - \rangle_f^*)\,,
\qquad\qquad  
C^f_4\ =\   \imag(\langle + \rangle_f \langle - \rangle_f^*)\,.
\end{eqnarray}
%
%
For our numerical results, we make the following choices of parameters:
\begin{eqnarray}
  \label{MSSM1}
&&\tan\beta=50, \ \ M_{H^\pm}^{\rm pole}=155~~{\rm GeV},
\nonumber \\
&&M_{\tilde{Q}_3} = M_{\tilde{U}_3} = M_{\tilde{D}_3} =
M_{\tilde{L}_3} = M_{\tilde{E}_3} = M_{\rm SUSY} = 0.5 ~~{\rm TeV},
\nonumber \\
&& |\mu|=0.5 ~~{\rm TeV}, \ \
|A_{t,b,\tau}|=1 ~~{\rm TeV},   \ \
|M_2|=|M_1|=0.3~~{\rm TeV}, \ \ |M_3|=1 ~~{\rm TeV},
\nonumber \\
&&
\Phi_\mu = 0^\circ, \ \
\Phi_A=\Phi_{A_t} = \Phi_{A_b} = \Phi_{A_\tau} = 90^\circ, \ \
\Phi_1 = \Phi_2 = 0^\circ,
\end{eqnarray}
and we consider two values for the phase of the gluino mass parameter
$M_3$: $\Phi_3 = -10^\circ\,, -90^\circ$.
For $\Phi_3 = -10^\circ$, {\tt CPsuperH}~\cite{CPsuperH} yields for
the masses and widths of the neutral Higgs bosons: \\
\begin{eqnarray}
&&
M_{H_1}=120.2~~{\rm GeV}, \ \
M_{H_2}=121.4~~{\rm GeV}, \ \
M_{H_3}=124.5~~{\rm GeV}, \ \
\nonumber \\  &&
\Gamma_{H_1}=1.19~~{\rm GeV}, \ \ \ \
\Gamma_{H_2}=3.42~~{\rm GeV}, \ \ \ \ \ \
\Gamma_{H_3}=3.20~~{\rm GeV},
\end{eqnarray}
and for $\Phi_3 = -90^\circ$:
\begin{eqnarray}
&&
M_{H_1}=118.4~~{\rm GeV}, \ \
M_{H_2}=119.0~~{\rm GeV}, \ \
M_{H_3}=122.5~~{\rm GeV}, \ \
\nonumber \\  &&
\Gamma_{H_1}=3.91~~{\rm GeV}, \ \ \ \
\Gamma_{H_2}=6.02~~{\rm GeV}, \ \ \ \ \ \
\Gamma_{H_3}=6.34~~{\rm GeV},
\end{eqnarray}
respectively.  In the above  MSSM scenario, which has originally  been
introduced in  \cite{ELP1,ELP2},  all the  three  neutral CP-violating
Higgs bosons are nearly degenerate, with masses of $\sim 120$~GeV, and
mix strongly through off-diagonal  absorptive self-energy effects.  We
call a scenario with  these properties, which  can only be realized in
a  CP-violating MSSM, the  {\em three-way  mixing} or
{\em tri-mixing} scenario.


For  $f=b$,  the polarization  of the  $b$  quark and  the  ${\bar b}$
anti-quark cannot  be  measured, and  the only  observable is the total
cross section, which is given by
\begin{equation}
M^2\frac{\partial^2\sigma_{\rm tot}^{b}}{\partial y \partial M^2}\ =\ 4 K
\hat{\sigma}_1^b\,
M^2\frac{\partial^2 {\cal L}}{\partial y\partial M^2}\; ,
\end{equation}
where we include    the   perturbative  QCD correction   $K     \equiv
1+\frac{\alpha_s(\hat{s})}{\pi}(\pi^2+11/2)$ \cite{KFACTOR}.  
The factor 4 comes from
the sum over the $b$- and $\bar{b}$-quark polarizations~\cite{CFLMP}.

\begin{figure}[htb]
\vspace{-1.5cm}
\centerline{\epsfig{figure=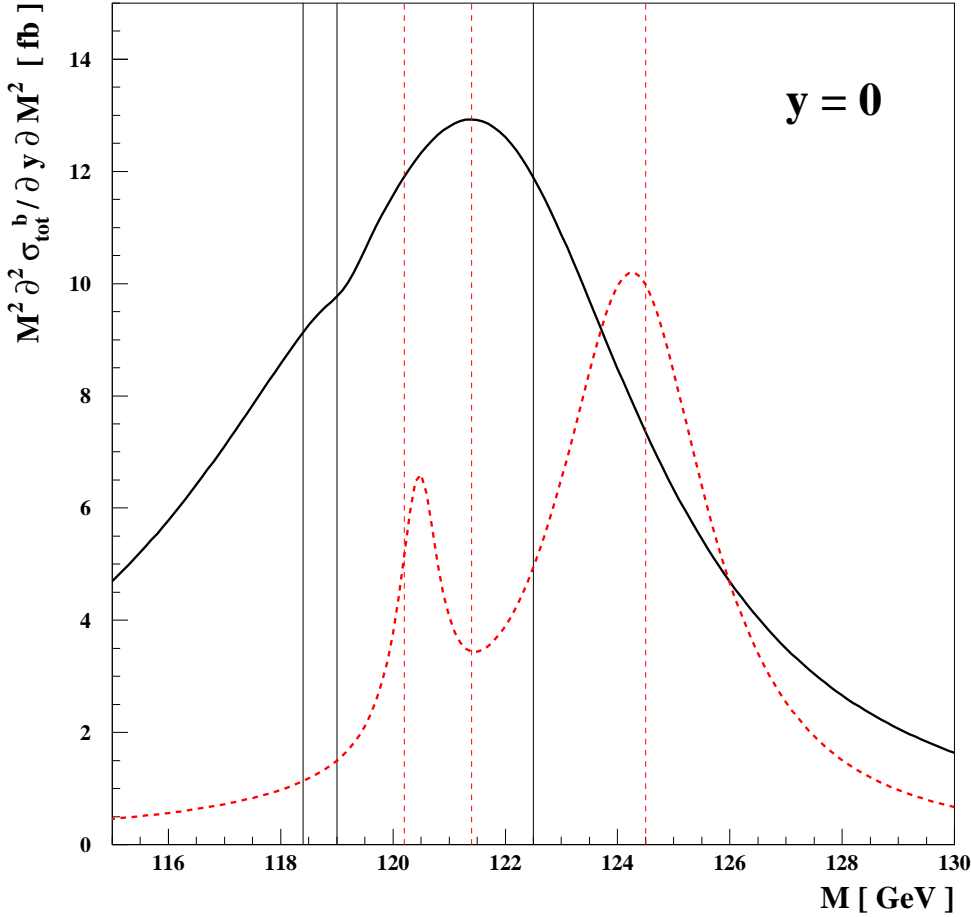,height=14cm,width=14cm}}
\vspace{-0.9 cm}
\caption{{\it     The      hadronic   level       cross        section
$M^2\frac{\partial^2\sigma_{\rm      tot}^{b}}{\partial y     \partial
M^2}\,(y=0)$ when  the produced Higgs  bosons  decay into $b$  quarks,
calculated using CTEQ6M  PDFs. Tri-mixing  scenarios have been   taken
with $\Phi_3=-90^\circ$ (solid  lines) and $\Phi_3=-10^\circ$  (dotted
lines).  The vertical lines  indicate the three Higgs-boson  pole-mass
positions.  }}
\label{fig:bquark}
\end{figure}

Figure~\ref{fig:bquark} displays  two numerical examples of 
three-way mixing
scenarios with $\Phi_3=-90^\circ$ (solid lines) and $\Phi_3=-10^\circ$
(dashed lines). In this case, we have taken CTEQ6M and $Q_{\rm min}=1$
GeV.  The differential cross section  becomes as large as $\sim 13$~fb
for   $\Phi_3=-90^\circ$, in which case   it   exhibits a single  peak
located between  the  pole masses,  that  are indicated  by the  solid
vertical  lines.  This peak is  broader than the expected missing-mass
resolution  $\delta M \sim 1$~GeV. On  the other hand,  twin peaks are
discernible   when $\Phi_3=-10^\circ$,  thanks   to  the expected good
resolution in the  Higgs mass: $\delta M  \sim 1$~GeV.  The twin peaks
appear close to (but not at) the  outer pair of pole masses, indicated
by the vertical dashed lines, and the  cross section actually exhibits
a dip at  the second pole mass.  These  examples  demonstrate that the
unique sensitivity  to CP-conserving observables  provided by the good
mass  resolution  in double-diffraction events  would  in turn provide
sensitivity to the CP-violating phase $\Phi_3$.

For $f=\tau$, we  have four observables which  can be constructed from
$\hat\sigma^\tau_i$    with $i=1-4$. In   particular,  to  analyze the
signatures     of  CP    violation       in   the    production     of
longitudinally-polarized $\tau$ leptons, we define
\begin{equation}
\Delta\sigma_{\rm CP}^{\tau}\ \equiv\
 \sigma(pp\rightarrow p H_i p \,;\, H_i \rightarrow \tau^+_R\tau^-_R)\:
-\: \sigma(pp\rightarrow p H_i p \,;\, H_i \rightarrow \tau^+_L\tau^-_L)\,.
\end{equation}
We then have the total and CP-violating cross sections given by
\begin{eqnarray}
M^2\frac{\partial^2\sigma_{\rm tot}^{\tau}}{\partial y \partial M^2} 
&=& 4 K \hat{\sigma}_1^\tau
M^2\frac{\partial^2 {\cal L}}{\partial y\partial M^2} \,,
\nonumber \\
M^2\frac{\partial^2\Delta\sigma_{\rm CP}^{\tau}}{\partial y \partial M^2} 
&=& 4 K \hat{\sigma}_2^\tau
M^2\frac{\partial^2 {\cal L}}{\partial y\partial M^2} \,.
\end{eqnarray}
Finally, we define the CP-violating asymmetry $a_{\rm CP}$ as
\begin{equation}
a_{\rm CP}^\tau\ \equiv\ \frac{ M^2\frac{\partial^2
\Delta\sigma_{\rm CP}^\tau}{\partial y \partial M^2}}{ 
M^2\frac{\partial^2\sigma_{\rm tot}^\tau}{\partial y \partial M^2}}\ 
=\ \frac{\hat\sigma^\tau_2}{\hat\sigma^\tau_1}\; .
\end{equation}

\begin{figure}[htb]
\vspace{-1.5cm}
\centerline{\epsfig{figure=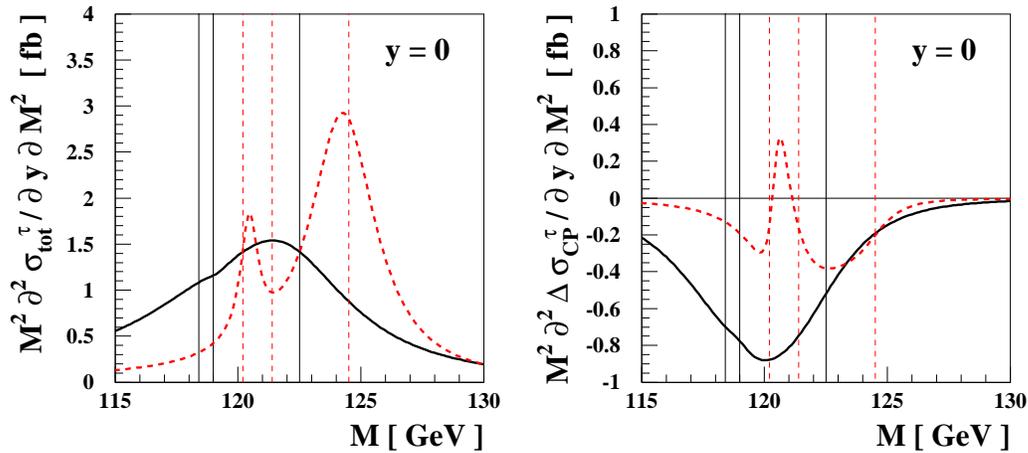,height=14cm,width=14cm}}
\vspace{-6.9 cm}
\caption{{\it  The  hadron-level  CP-conserving and CP-violating cross
sections   when    Higgs   bosons   decay    into   $\tau$    leptons:
$M^2\frac{\partial^2\sigma_{\rm tot}^{\tau}}{\partial y \partial M^2}$
(left)  and $M^2\frac{\partial^2\Delta\sigma_{\rm CP}^{\tau}}{\partial
y   \partial  M^2}$ (right), calculated using   CTEQ6M  PDFs.  We have
considered  tri-mixing scenarios with $\Phi_3=-90^\circ$ (solid lines)
and $\Phi_3=-10^\circ$ (dotted lines). The vertical lines indicate the
three Higgs-boson pole-mass positions.  }}
\label{fig:tau}
\end{figure}

Figure~\ref{fig:tau}   displays  numerical   examples   in  tri-mixing
scenarios with $\Phi_3=-90^\circ$ (solid lines) and $\Phi_3=-10^\circ$
(dotted lines).  We  have again taken CTEQ6M and  $Q_{\rm min}=1$ GeV,
and find cross sections  as large as $\sim 1.5 - 3$  fb. The peaks and
dips  in the  total cross  section are  located relative  to  the pole
masses in  the same  way as for  the ${\bar  b} b$ final  state, again
offering sensitivity  to the CP-violating  phase.  Moreover, comparing
Figs.~\ref{fig:bquark}  and \ref{fig:tau},  we see  that  the relative
sizes of the peaks are different  for the two values of $\Phi_3$, thus
providing more sensitivity to this CP-violating parameter.

\bigskip\bigskip

\begin{figure}[htb]
\vspace{-1.5cm}
\centerline{\epsfig{figure=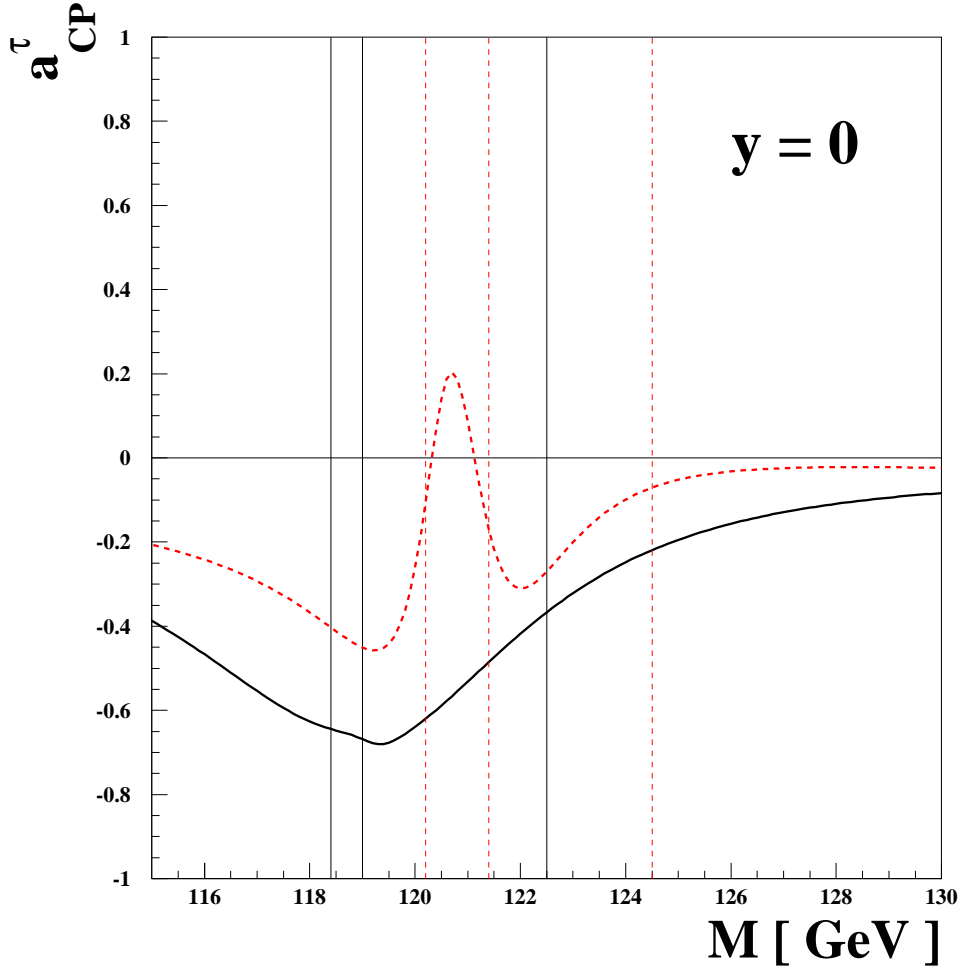,height=14cm,width=14cm}}
\vspace{-0.5 cm}
\caption{{\it 
The CP-violating asymmetry $a^\tau_{\rm CP}$ observable in tri-mixing 
scenarios when Higgs bosons 
decay into $\tau$ leptons. The line styles are the same as in 
Fig.~\ref{fig:tau}.
}}
\label{fig:acp}
\end{figure}

The CP-violating  cross-section difference  observable in the  $\tau^+
\tau^-$ final state is shown in the right panel of Fig.~\ref{fig:tau}.
In the case $\Phi_3 =  -90^\circ$, the cross-section difference always
has the same sign and is maximized between the pole masses, whereas in
the case $\Phi_3 = -10^\circ$ it is generally smaller and exhibits two
sign  changes.   The expected missing-mass   resolution $\delta M \sim
1$~GeV should be  sufficient to resolve  some of these structures.  We
expect that a CP asymmetry $a_{\rm CP}^\tau$ larger than  10 \% may be
detected with an integrated luminosity  $\gsim 100$~fb$^{-1}$.  As  we
see   in  Fig.~\ref{fig:acp},   the CP-violating  observable   $a_{\rm
CP}^\tau$  attains values  considerably  larger than 10\%  in both the
scenarios studied.  Whereas the  individual  cross sections depend  on
the PDFs  used,  the  CP   asymmetry shown   in  Fig.~\ref{fig:acp} is
insensitive to this choice.

\section{Conclusions}
\label{sec:conclusion}

We  have extended  our  previous studies  of  CP-violating MSSM  Higgs
scenarios   with  large  tri-mixing~\cite{ELP1,ELP2}   to  diffractive
production  at  the LHC.   The  production  cross  sections are  large
compared  to those  in the  Standard  Model, and  the good  Higgs-mass
resolution obtainable via the missing-mass method should enable one to
disentangle  the different  adjacent resonant  peaks.  Although  it is
difficult to  construct CP-violating observables in $H_i  \to {\bar b}
b$ decays  without tagging the final  protons~\cite{KMR} or analyzing
the $b$-quark  decay products, observations  of $\tau^\pm$ helicities
would permit a CP asymmetry to  be measured in $H_i \to \tau^+ \tau^-$
decays.   This information  may also  be used  to further  resolve the
underlying resonant  dynamics of a strongly  mixed Higgs-boson system,
thereby offering a sensitive window  into CP violation and new physics
due to an extended Higgs sector, such as the MSSM Higgs sector.

This example shows that exclusive  double diffraction may offer unique
possibilities  for exploring Higgs    physics in ways  that  would  be
difficult or even  impossible   in  inclusive  Higgs  production.   In
particular,   we  have   shown   that  exclusive    double diffraction
constitutes an  efficient CP and   lineshape analyzer of  the resonant
Higgs-boson dynamics in multi-Higgs  models.  In the specific  case of
CP-violating MSSM Higgs physics discussed  here, which is  potentially
of  great  importance    for electroweak  baryogenesis,    diffractive
production may be the most promising probe at the LHC.

\subsection*{Acknowledgements}
We thank  Jeff Forshaw for helpful discussions  and valuable comments.
The work  of JSL  and AP is  supported in  part by the  PPARC research
grant PPA/G/O/2002/00471.

\newpage


\clearpage
\noindent

\end{document}